\documentclass[a4paper,11pt]{article}
\usepackage{pos}

\usepackage{graphicx}
\usepackage{subcaption}

\title{Nuclear Transition in the Strong Coupling Limit}

\author[a]{Jangho Kim}
\author*[b]{Pratitee Pattanaik}
\author[b]{Wolfgang Unger}

\affiliation[a]{Institute for Advanced Simulation (IAS-4), Forschungszentrum Jülich\\
Wilhelm-Johnen-Straße, 52428 Jülich, Germany}

\affiliation[b]{Fakult$\ddot{a}$t f$\ddot{u}$r Physik, Universit$\ddot{a}$t Bielefeld,\\
Universit$\ddot{a}$tstrasse 25, D33619 Bielefeld, Germany}

\emailAdd{j.kim@fz-juelich.de}
\emailAdd{pratiteep@physik.uni-bielefeld.de}
\emailAdd{wunger@physik.uni-bielefeld.de}

\abstract{Lattice QCD at finite baryon chemical potential has the infamous sign problem which hinders Monte Carlo simulations. This can be remedied by a dual representation that makes the sign problem mild. In the strong coupling limit, the dual formulation with staggered quarks is well established. We have used this formulation to study the quark mass dependence of the baryon mass and the nuclear transition. This allows us to quantify the nuclear interaction. We have also compared our Monte Carlo results with mean field predictions.}

\FullConference{%
  39th International Symposium on Lattice Field Theory-Lattice2022,\\
  8-13 August 2022\\
  Bonn, Germany
}

\begin{document}
\maketitle

\section{Introduction}

It is known from experiments that at low temperatures, there is a phase transition between dilute hadron gas and dense nuclear matter as the baryon chemical potential increases. This transition is of first order and terminates at about $T_c=16$ MeV in a critical end point. It is interesting to study the dependence of the nuclear force and the phase transition on different parameters \cite{Epelbaum:2008ga}. \\

The nuclear force between baryons to form nuclear matter is due to the residual strong interactions between quarks and gluons. Therefore, it should be accurately described by QCD \cite{PhysRevLett.99.022001}. However, at finite baryon density, lattice QCD has the infamous sign problem which does not allow us to perform direct Monte Carlo simulations on the lattice. Various methods have been proposed to overcome this, but work only for small lattice volumes and chemical potentials. An alternative established method is to study lattice QCD using the strong coupling expansion. \\

We study lattice QCD in the strong coupling limit (infinite bare gauge coupling, $\beta$ = 0) which is far from the continuum limit. In this limit, the gauge integration is performed first, followed by the Grassmann integration to get a dual formulation \cite{ROSSI1984105}. The sign problem is mild and can be taken into account exactly rather than probabilistically in this formulation. The new dual degrees of freedom are color-singlet mesons and baryons. Apart from this, chiral symmetry breaking and confinement properties of QCD are still present in the strong coupling limit. The complete phase diagram can be determined in this limit, making this an interesting effective theory.\\

Strong coupling lattice QCD does not have a well defined lattice spacing. Therefore, we need to scale the observables with baryon mass to have dimensionless quantities that can be compared with other theories. We will discuss the dependence of baryon mass on quark mass. The nuclear transition and interaction using the dual formulation is also discussed. Another effective approach is the Polyakov loop effective theory based on the hopping parameter expansion \cite{Langelage_2014}. However, this is limited to rather large quark masses. Therefore, to establish a region of overlap, we have chosen to simulate the Dual formulation also at large quark masses. Lastly, we have also compared the results with the mean field theory approach.

\section{Dual formulation}

In the strong coupling limit, there is no Wilson plaquette action $S_G$ since $\beta$=0. Hence, the link integral of the QCD partition function factorizes and we can rewrite it as a product of one-link integrals:
\begin{align}
Z &=\int \prod_{x}\left(\mathrm{~d} \chi_{x} \mathrm{~d} \bar{\chi}_{x} \mathrm{e}^{2 a m_{q} \bar{\chi}_{x} \chi_{x}} \textcolor{red}{\prod_{\hat{\mu}}\left[\mathrm{d} U_{\hat{\mu}}(x) \mathrm{e}^{\eta_{\hat{\mu}}(x)\left(\bar{\chi}_{x} U_{\hat{\mu}}(x) \chi_{x+\hat{\mu}}-\bar{\chi}_{x+\hat{\mu}} U_{\hat{\mu}}^{\dagger}(x) \chi_{x}\right)}\right]}\right) \nonumber \\
&=: \quad \int \prod_{x}\left(\mathrm{~d} \chi_{x} \mathrm{~d}_{\chi_{x}} \mathrm{e}^{2 a m_{q} \bar{\chi}_{x} \chi_{x}} \textcolor{red}{\prod_{\hat{\mu}} z(x, \hat{\mu})}\right)
\end{align}

$z(x,\hat{\mu})$ is the one-link gauge integral of individual links. This one-link group integral can be solved and rewritten in terms of new variables:
\begin{align}
M(z)=\bar{\chi}_{z} \chi_{z} \quad \text { and } \quad B(z)=\frac{1}{N !} \varepsilon_{i_{1} \ldots i_{N}} \chi_{z, i_{1} \cdots \chi_{z, i_{N}}}
\end{align}

These are the mesonic and baryonic color singlet states respectively. The solution of the one-link integral has terms like $(M(x)M(y))^k$ (dimers), $\bar{B}(y)B(x)$ and $\bar{B}(x)B(y)$ (baryon links) in term of the new variables. After the gauge integration, we can perform the remaining Grassmann integral which gives us terms of the form $(2am_qM(x))^{n_x}$ (monomers).\\

The grassmann constraints needs to be satisfied at each lattice site to get non-zero results. This leads to baryons forming only self avoiding loops ($\ell$) and a constraint on the number of monomers, dimers and baryon loops per site:
\begin{align}
  n_x+\sum_{\pm \hat{\mu}} \left( k_{x\hat{\mu}} + \frac{N_c}{2} |l_{x\hat{\mu}}| \right ) = N_c
 \end{align}

The partition function in terms of the dual variables ($n_x$,$k_b$,$\ell$) for $N_c=3$ can be written as 
 \begin{align}
  Z=\sum_{\{k, n, \ell\}} \textcolor{red}{\prod_{b=(x, \hat{\mu})} \frac{\left(3-k_{b}\right) !}{3 ! k_{b} !} \gamma^{2 k_{b} \delta_{0, \hat{\mu}}}} \textcolor{blue} {\prod_{x} \frac{3 !}{n_{x} !}\left(2 a m_{q}\right)^{n_{x}}} \textcolor{black}{\prod_{\ell} w(\ell)}
 \end{align}

The term in red are contribution of the dimers, the term in blue are due to the monomers and the term $w(\ell)$ is the contribution from the baryon loops given by 
 \begin{align}
  w(\ell)=\frac{1}{\prod_{x \in \ell} 3 !}\sigma(\ell) \gamma^{3 N_{\hat{0}}} \exp \left(3 N_{t} r_{\ell} a_{t} \mu\right)
 \end{align}
 
 where $\sigma(l)$ is the sign of a configuration which can be explicitly calculated. The dual formulation has a +1 or -1 sign, therefore sign reweighting techniques can be applied. The dual partition function is simulated using the worm algorithm.\\
 
 Observables in terms of the new degrees of freedom can be calculated and we are interested in the energy density which is given by 
 \begin{align}
 a^{4} \epsilon= \frac{\xi}{\gamma} \frac{\partial \gamma}{\partial \xi}\left\langle 2 N_{D t}+3 N_{B t}\right\rangle - \left\langle N_M \right\rangle
\end{align}

where $N_{Dt}$ is the number of temporal dimers, $N_{Bt}$ is the number of temporal baryon segments and $N_M$ is the number of monomers in a configuration, $\gamma$ is the bare anisotropy and $\xi=a/a_t$ is the physical anisotropy. The relation between $\xi$ and $\gamma$ was determined and it is seen that even for isotropic lattices where $\gamma=1$, $\frac{\partial \gamma}{\partial \xi}|_{\gamma=1}$ is non-trivial and depends on the quark mass.

\section{Baryon mass and nuclear interaction}

A static baryon in the dual formulation is a closed baryon loop in the temporal direction as shown in Fig. 1a. The loop is built of links $\bar{B}(x)B(y)$ which is a baryon hopping (annhilation of a baryon at site $x$ and creation at site $y$). At low temperatures ($aT=\frac{1}{N_t}$), the probability for this hopping is proportional to $e^{-am_B}$ where $m_B$ is the mass of the baryon. Therefore, for a static baryon having $N_t$ links, the probability is proportional to $e^{-am_BN_t} = e^{\frac{-m_B}{T}}$. \\

\begin{figure}[]
  \centering
  \begin{subfigure}[b]{0.4\linewidth}
    \centering
    \includegraphics[height=3.5cm]{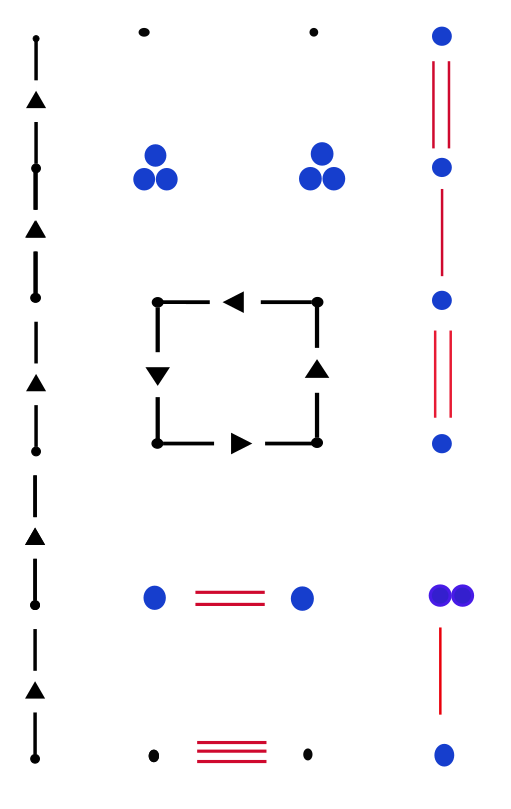}
     \caption{With a static baryon}
  \end{subfigure}
  \hspace{3em}
  \begin{subfigure}[b]{0.4\linewidth}
  \centering
    \includegraphics[height=3.5cm]{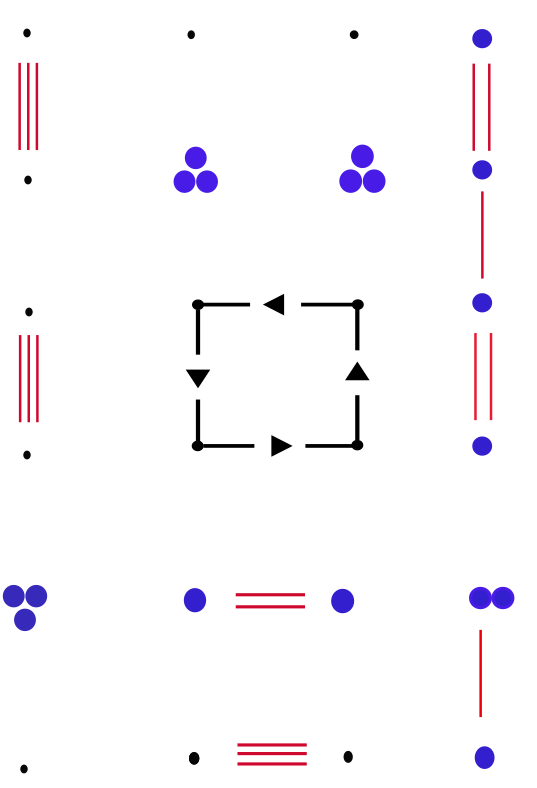}
    \caption{Without a static baryon}
  \end{subfigure}
  \caption{Configurations on a 1+1$D$ lattice. The vertical direction is the temporal direction.}
\end{figure}

However, this probability can be expressed as $e^{\frac{-\Delta F}{T}}$ where $\Delta F$ is the difference in free energy between a configuration having a static baryon world line (Fig. 1a) and a configuration without (Fig. 1b). Therefore, equating the two expressions for the probability, we get $m_B=\Delta F$. This has been used to calculate the baryon mass before \cite{Fromm:2010lga}. At low temperatures, the free energy ($F=E-TS$) is approximately equal to the energy. So baryon mass can be calculated by $\Delta E$ instead. We have used a lattice of $8^3 \times 8$ at a temperature $0.125$ to perform the simulations. Fig. 2a shows the baryon mass at different quark masses. We find that the baryon mass increases as the quark mass increases.\\

\begin{figure}[]
  \centering
  \begin{subfigure}[b]{0.45\linewidth}
    \centering
    \includegraphics[height=4cm]{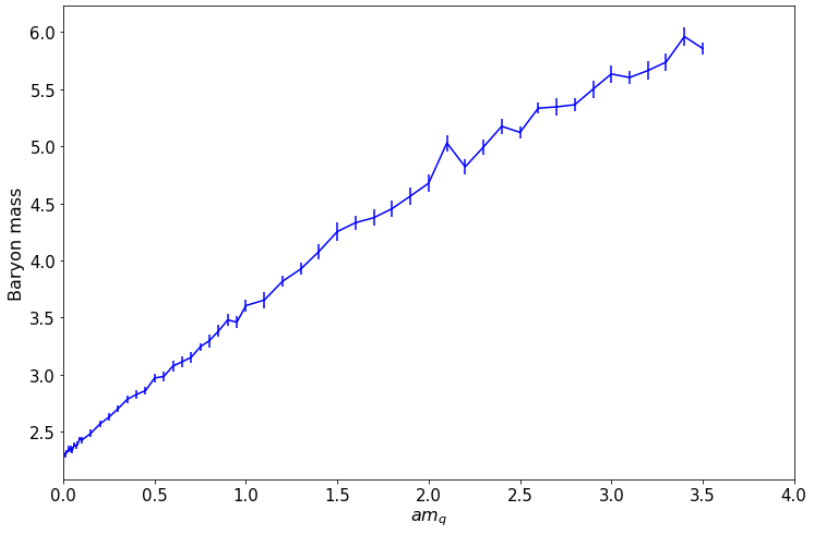}
     \caption{Baryon mass at different quark masses}
  \end{subfigure}
  \hspace{1em}
  \begin{subfigure}[b]{0.45\linewidth}
  \centering
    \includegraphics[height=4cm]{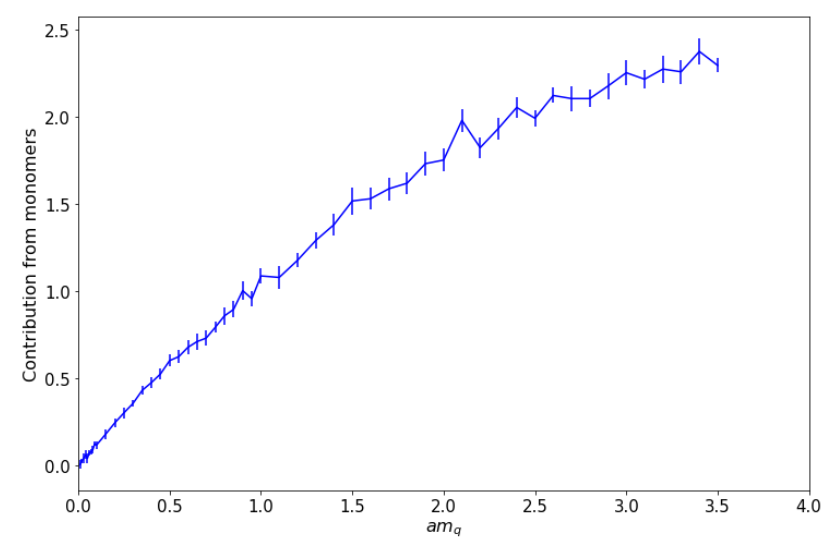}
    \caption{Contribution to baryon mass from monomers}
  \end{subfigure}
  \newline
  \centering
  \begin{subfigure}[b]{0.45\linewidth}
    \centering
    \includegraphics[height=4cm]{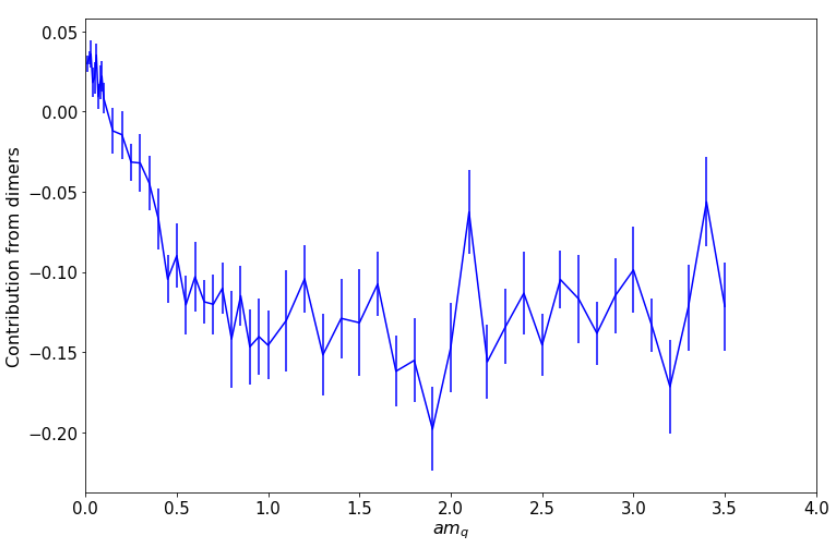}
     \caption{Contribution to baryon mass from dimers}
  \end{subfigure}
  \hspace{1em}
  \begin{subfigure}[b]{0.45\linewidth}
    \centering
    \includegraphics[height=4cm]{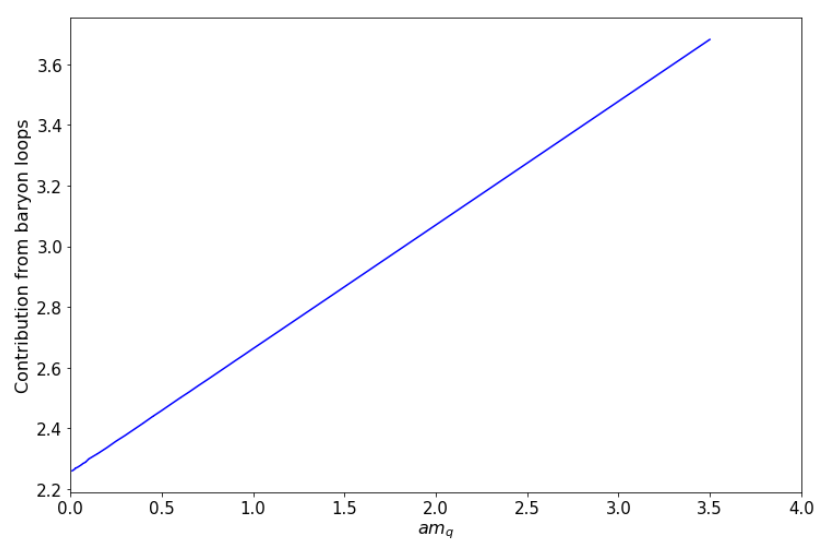}
     \caption{Contribution to baryon mass from baryon loops}
  \end{subfigure}
  \caption{Results on $8^3\times 8$ lattice at $aT$=0.125}
\end{figure}

Due to the dual formulation, we can also see the contributions of different degrees of freedom on the baryon mass. Fig. 2b shows the contribution from monomers. We find that it increases as the quark mass increases. Fig. 2c shows the contribution of dimers to the baryon mass. The contribution is small compared to the contribution from monomers and baryon loops. Fig. 2d shows the contribution of baryon loops to the quark mass.\\

The nuclear transition for low quark masses at intermediate temperatures is established at the strong coupling limit \cite{https://doi.org/10.48550/arxiv.1611.09120}. We have extended the results to higher quark masses and lower temperatures where the nuclear critical end point moves to large $\mu_B/T$.\\

With the dual formulation, we can calculate the baryon density and the baryon susceptibility at different baryon chemical potential for different quark masses. We know that for a first order transition, the baryon density shows a discontinuity. For larger quark masses at a fixed temperature, this transition becomes continuous which is a crossover as seen in Fig. 3a, and the baryon susceptibility has a peak which does not diverge (Fig. 3b). If the transition is a crossover then there is no well defined value for $\mu_{Bc}$, so we take the critical chemical potential to be the value at which the baryon density is $n_B=0.5$. We also know that a first order transition scales with the spatial lattice volume whereas a crossover transition does not. By performing simulations at lattice volumes of $4^3\times 8$, $6^3\times 8$ and $8^3\times 8$ we aim to narrow down the critical end point for fixed temperature $aT=0.125$. We find that the first order transition has a smaller gap beyond quark mass $am_q=1.5$ and it has vanished at $am_q=1.8$. The best guess for the nuclear critical end point is $am_q^c=1.7(1)$, see Fig. 4a.\\

\begin{figure}[]
  \centering
  \begin{subfigure}[b]{0.45\linewidth}
    \centering
    \includegraphics[height=4cm]{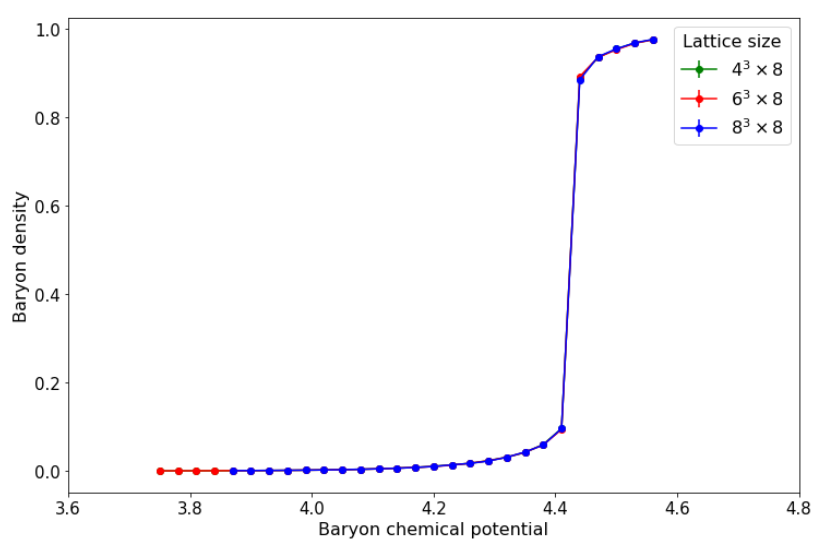}
     \caption{Baryon density for $am_q=1.9$}
  \end{subfigure}
  \hspace{1em}
  \begin{subfigure}[b]{0.45\linewidth}
  \centering
    \includegraphics[height=4cm]{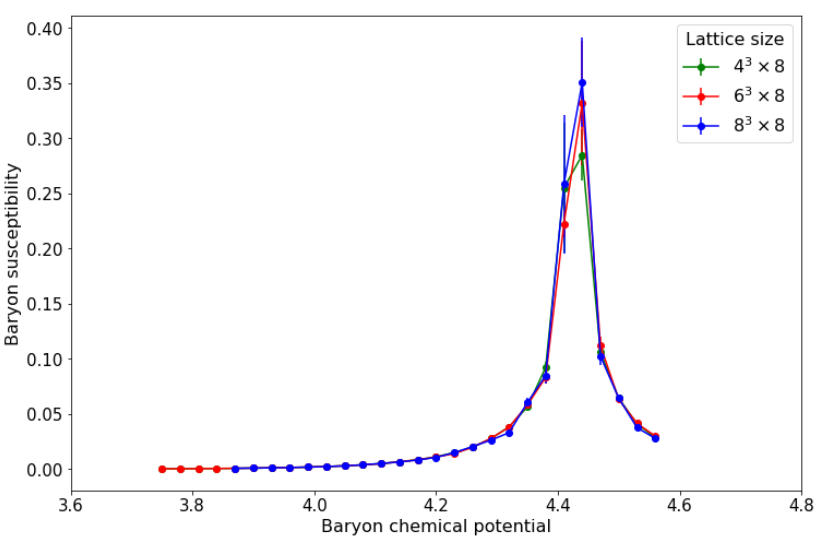}
    \caption{Baryon susceptibility for $am_q=1.9$}
  \end{subfigure}
  \caption{Left: Baryon density for $am_q=1.9$. Right: Baryon susceptibility for $am_q=1.9$. The peak of baryon susceptibility does not change with varying lattice volumes implying a crossover.}
\end{figure}

We would like to mention that for higher quark masses, the autocorrelations between the configurations is high which makes it challenging to perform the simulations. We have overcome this by using static updates with weights from 1-dim QCD to equilibrate between the low density phase dominated by monomers and the high density phase dominated by baryons.\\

\begin{figure}[]
  \centering
  \begin{subfigure}[b]{0.45\linewidth}
    \centering
    \includegraphics[height=4cm]{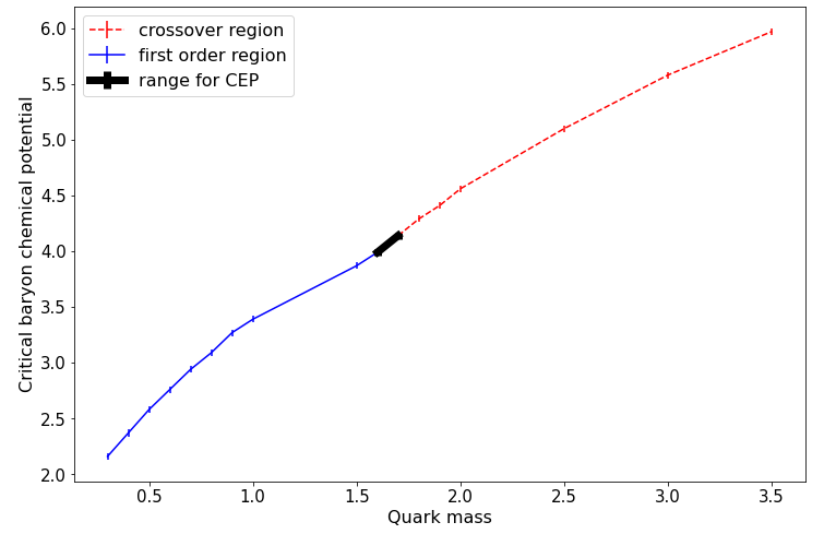}
     \caption{$\mu_{Bc}$ for different $am_q$}
  \end{subfigure}
  \hspace{1em}
  \begin{subfigure}[b]{0.45\linewidth}
  \centering
    \includegraphics[height=4cm]{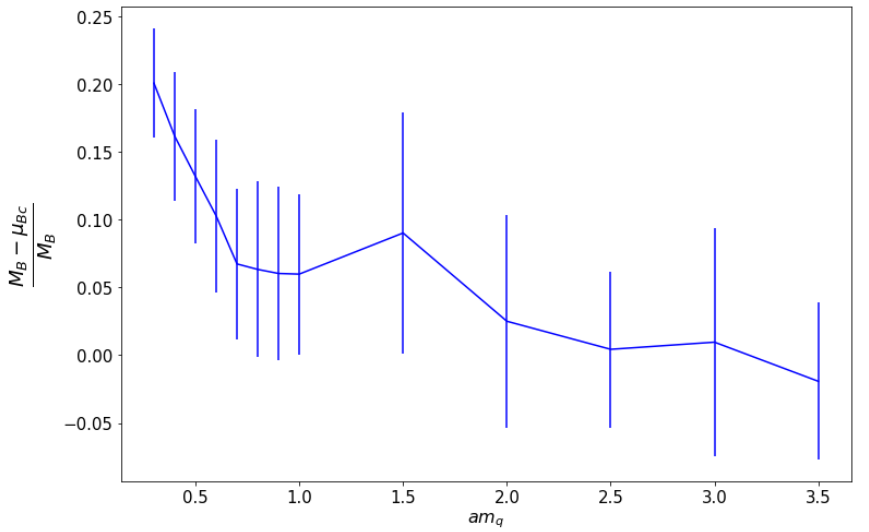}
    \caption{Nuclear interaction for different $am_q$}
  \end{subfigure}
  \caption{Left: The critical baryon chemical potential for different quark masses. The blue line shows the first order region and the red line shows the crossover region. The range for critical end point is marked in black. Right: Nuclear interaction scaled with the baryon mass. It goes to zero as the quark mass increases.}
\end{figure}

The change in free energy to have a non-interacting baryon in a dilute bath of mesons is given by its mass. This happens at the transition where we get a non-zero baryon density. Therefore, the critical baryon chemical potential should be roughly the mass of the baryon. Due to nuclear interactions, it is seen that there is a difference between $\mu_{B,c}$ and $m_B$ which we can quantify \cite{PhysRevD.45.3228}.\\

The difference between the critical baryon chemical potential and the baryon mass, scaled with the baryon mass at different quark masses, is plotted in Fig. 4b. This quantifies the nuclear interaction and we see that it decreases as the quark mass increases. It is expected that as the quark mass tends to infinity, the nuclear interaction tends to zero which is consistent with our results. Higher statistics are needed to get results with less error.

\section{Comparision with mean field theory}

We can compare the Monte Carlo results with the mean field approach of LQCD which is based on a $\frac{1}{d}$ expansion ($d$ is the spatial dimension) of the partition function followed by Hubbard-Stratonovich transformation \cite{PhysRevD.69.094501}. The free energy obtained after this procedure as a function of the chiral condensate ($\sigma$), temperature ($T$) and chemical potential ($\mu_B$) is given by

\begin{align}
   F_{\text {eff }}\left[\sigma, T, \mu_{\mathrm{B}}\right]&=\frac{N_{\mathrm{c}} d}{4}\sigma^{2}-T \log \left\{2 \cosh \left[\mu_{\mathrm{B}} / T\right]+\frac{\sinh \left[\left(N_{\mathrm{c}}+1\right) E / T\right]}{\sinh [E / T]}\right\} \nonumber \\
   E[m]&=\text{arcsinh}\left[\sqrt{m^2+\left(\frac{d\sigma}{2}\right)^2+md\sigma}\right]
  \end{align}
  
$E[m]$ is the one-dimensional quark excitation energy which depends on the quark mass $m$. The global minimum of the free energy with respect to the chiral condensate gives the chiral condensate as a function of temperature, chemical potential and quark mass. At a given temperature and quark mass, the chiral condensate can be plotted as a function of the chemical potential and the chemical potential where there is a discontinuity in the chiral condensate indicates the critical chemical potential.\\

Using this, we have plotted the phase diagram for different quark masses in Fig 5a. We have plotted the critical baryon chemical potential obtained from mean field theory at $aT$=0.125 in Fig. 5b (orange). We have also plotted the critical baryon chemical potential from Monte Carlo simulations at $aT$=0.125 in Fig. 5b (green). We see that the mean field theory results and the dual formulation results have a small discrepancy. The critical baryon chemical potential from 1-dim.~QCD is also plotted in Fig. 5b (black). This gives us a lower bound to both the baryon mass and the critical baryon chemical potential. We see that the Monte Carlo results approach the 1-dim.~QCD results as the quark mass becomes large.\\

The baryon mass $m_B$ can also be computed with the mean field theory \cite{KLUBERGSTERN1983447} from the equation: 
\begin{align}
\sinh \, (am_B) = \frac{1}{2} \left(m+\sqrt{m^2+1}\right)^{N} (2d)^{N/2}
\end{align}
This is plotted in Fig. 5b (red). This gives an upper bound to the baryon mass. The baryon mass from the dual formulation is plotted in Fig. 5b (blue). We see that it lies within the upper and lower bounds discussed.\\

For comparision, we have also plotted the baryon mass from hopping parameter expansion with Wilson fermions \cite{HOEK1982495} in Fig. 5b (brown). We see that at large quark masses, the dual formulation results are similar to this.

\begin{figure}[]
  \centering
  \begin{subfigure}[b]{0.45\linewidth}
    \centering
    \includegraphics[height=4cm]{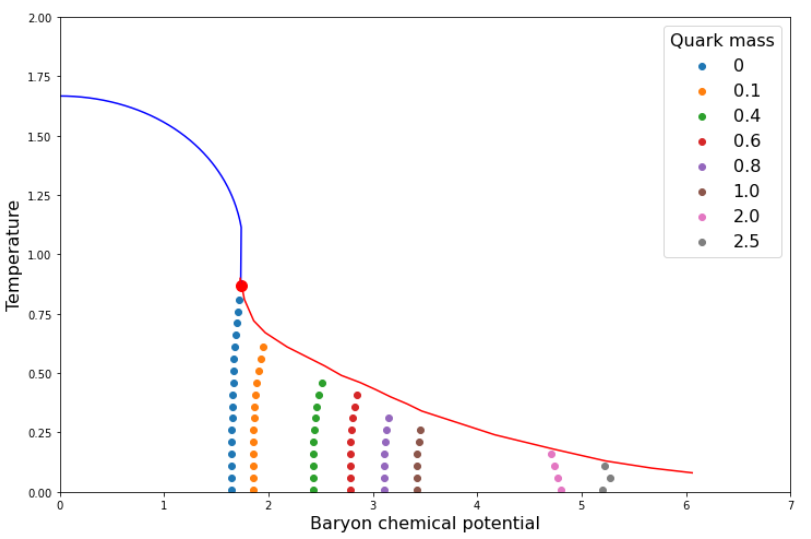}
     \caption{Phase diagram for different $am_q$}
  \end{subfigure}
  \hspace{1em}
  \begin{subfigure}[b]{0.45\linewidth}
  \centering
    \includegraphics[height=4cm]{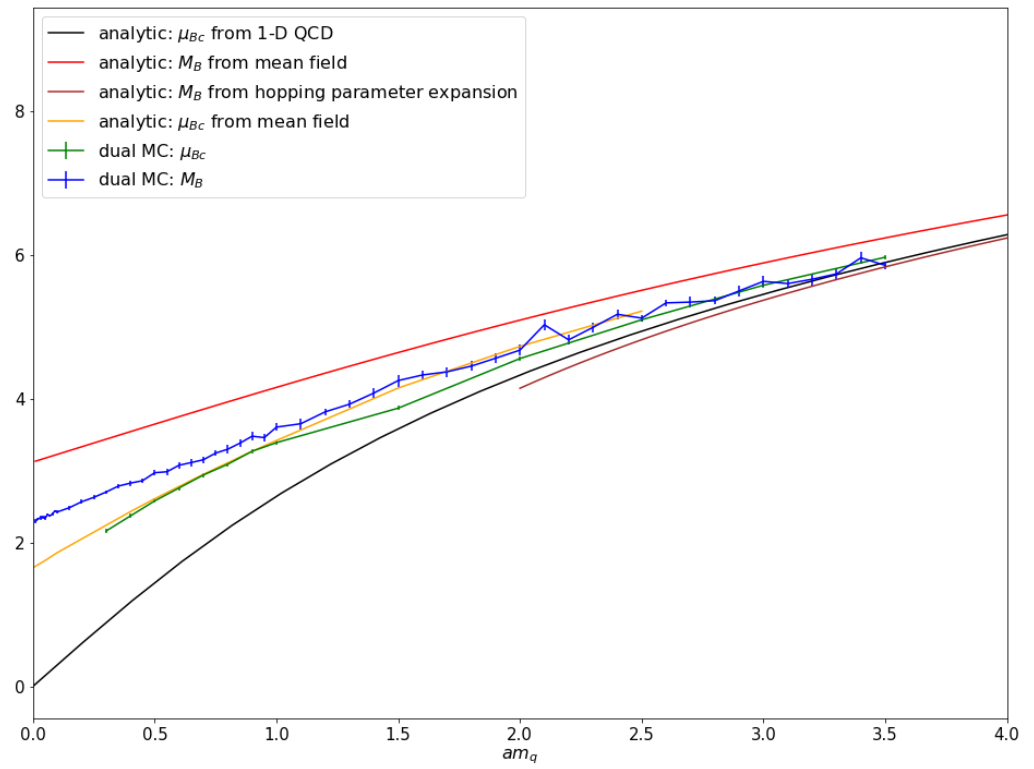}
    \caption{Baryon mass and $\mu_{Bc}$ for different $am_q$}
  \end{subfigure}
  \caption{Left: The phase diagram obtained from mean field theory. The blue line shows the second order transition at chiral limit. The dotted lines show the phase transition for different quark masses. The red line shows the critical end point for different quark masses. Right: The baryon mass from dual formulation, mean field theory and hopping parameter expansion with Wilson fermions is plotted. The critical chemical potential from dual formulation, mean field theory and 1-dim QCD is also plotted.}
\end{figure}

\section{Conclusion and Outlook}

We find that the dual formulation of lattice QCD has a mild sign problem, thus allowing for calculations at finite baryon density. The baryon mass, nuclear transition and nuclear interaction was found for different quark masses at $aT$=0.125. The nuclear interaction tends to zero when the quark mass is large. The results found from dual formulation have a small discrepancy with the mean field theory predictions. At large quark masses, the dual formulation results agree with the 3-dim.~effective theory \cite{Langelage_2014} based on hopping parameter expansion. \\

We plan to find the baryon mass and the nuclear interaction after including the gauge corrections in the Dual formulation. This is already done for small quark masses and we plan to extend it for large quark masses. We also plan to extend the simulations to two-flavor QCD using the continuous time limit.

\section{Acknowledgements}

This work was supported by the Deutsche Forschungsgemeinschaft (DFG, German Research Foundation) through the CRC-TR 211 ’Strong-interaction matter under extreme conditions’ - project number 315477589 - TRR 211. The authors gratefully acknowledge the funding of this project by computing time provided by the Paderborn Center for Parallel Computing (PC2). J.Kim was supported in part by the NSFC and the Deutsche Forschungsgemeinschaft (DFG, German Research Foundation) through the funds provided to the Sino-German Collaborative Research Center TRR110 "Symmetries and the Emergence of Structure in QCD" (NSFC Grant No. 12070131001, DFG Project-ID 196253076 - TRR 110).

\renewcommand*{\bibfont}{\small}
\bibliographystyle{unsrt}
\bibliography{main}

\end{document}